\documentclass[useAMS,usenatbib]{mn2e}

\usepackage{epsfig}
\usepackage{amsmath}
\usepackage{amssymb}
\usepackage{natbib}
\usepackage{threeparttable} 
\usepackage{booktabs}
\usepackage{float}
\usepackage{times}
\usepackage[normalem]{ulem}

\usepackage{hyperref} 
\hypersetup{
     colorlinks=true,
     linkcolor=purple,
     filecolor=purple,
     citecolor = blue,      
     urlcolor=cyan,
     }
     
\newcommand{\chan}{\textit{Chandra}}
\newcommand{\swift}{\textit{Swift}}
\newcommand{\rxte}{\textit{RXTE}}

\newcommand{\maxi}{\textit{MAXI}}

\newcommand{\gaia}{\textit{Gaia}}

\newcommand{\nicer}{\textit{NICER}}

\newcommand{\cnts}{\mathrm{c~s}^{-1}}

\newcommand{\mdotg}{\mathrm{g~s}^{-1}}

\newcommand{\xte}{XTE J1701--462}

\newcommand{\ks}{KS~1731--260}
\newcommand{\mxb}{MXB~1659--29}

\newcommand{\sax}{SAX J1808.4--3658}

\newcommand{\be}{\begin{equation}}
\newcommand{\ee}{\end{equation}}
\newcommand{\ba}{\begin{eqnarray}}
\newcommand{\ea}{\end{eqnarray}}

\newcommand{\eq}[1]{Equation~\eqref{#1}}
\newcommand{\fig}[1]{Figure~\ref{#1}}

\newcommand{\sect}[1]{Section~\ref{#1}}
\newcommand{\app}[1]{Appendix~\ref{#1}}

\newcommand{\Eq}[1]{Equation~\eqref{#1}}

\newcommand{\gcc}{\mbox{$\mathrm{g \, cm}^{-3}$}}
\newcommand{\Msun}{\mbox{$\mathrm{M}_\odot \,$}}

\newcommand{\HETE}{\mbox{HETE J1900.1--2455}}

\def \mnras {MNRAS}
\def \apj {ApJ}
\def \apjs {ApJS}
\def \apjl {ApJL}
\def \aap {A\&A}

\def\araa{Annu. Rev. Astron. Astrophys.}
\def \atel {Astron. Tel}

\def \pasj {PASJ}

\def \sovast {Soviet Astronomy}

\def \arnps {Annu. Rev. Nucl. \& Part. Sci.} 
\def \pr {Phys. Rev.}
\def \prc {Phys. Rev. C}
\def \prd {Phys. Rev. D}
\def \pre {Phys. Rev. E}
\def \prl {Phys. Rev. Lett.}

\def \nphysa {Nucl. Phys. A}
\def \nphysb {Nucl. Phys. A}
\def \physrep {Phys. Rep.}

\usepackage[usenames]{xcolor}

\title[The neutron star core of \HETE]{Constraining the properties of dense neutron star cores: \\ the case of the  low-mass X-ray binary \HETE}
\author[N. Degenaar et al.]
{N. Degenaar$^1$\thanks{e-mail: degenaar@uva.nl}, 
D.~Page$^{2}$, 
J.~van~den~Eijnden$^{1,3}$
M.~V.~Beznogov$^{2}$, 
R. Wijnands$^{1}$, 
\newauthor
M.~Reynolds$^{4}$ 
\\
$^1$Anton Pannekoek Institute for Astronomy, University of Amsterdam, Postbus 94249, 1090 GE Amsterdam, The Netherlands\\
$^2$Instituto de Astronom\'ia, Universidad Nacional Aut\'onoma de M\'exico, Ciudad Universitaria, Ciudad de M\'exico, CDMX 04510, Mexico\\
$^3$Astrophysics, Department of Physics, University of Oxford, Denys Wilkinson Building, Keble Road, Oxford OX1 3RH, UK \\
$^4$Department of Astronomy, University of Michigan, 1085 South University Avenue, Ann Arbor, MI 48109, USA
}

\begin{document}

\date{Accepted 2021 July 23. Received 2021 July 22; in original form 2021 June 16}

\pagerange{\pageref{firstpage}--\pageref{lastpage}} \pubyear{0000}

\maketitle

\label{firstpage}

\begin{abstract}
Measuring the time evolution of the effective surface temperature of neutron stars can provide invaluable information on the properties of their dense cores. Here, we report on a new \chan\ observation of the transient neutron star low-mass X-ray binary \HETE, which was obtained $\approx$2.5~yr after the end of its $\approx$10-yr long accretion outburst. The source is barely detected during the observation, collecting only six net photons, all below 2 keV. Assuming that the spectrum is shaped as a neutron star atmosphere model, we perform a statistical analysis to determine a $1\sigma$ confidence upper range for the neutron star temperature of $\approx$30--39~eV (for an observer at infinity), depending on its mass, radius, and distance. Given the heat injected into the neutron star during the accretion outburst, estimated from data provided by all-sky monitors,
the inferred very low temperature suggests that the core either has a very high heat capacity or undergoes very rapid neutrino cooling. 
While the present data do not allow us to disentangle these two possibilities, both suggest that a significant fraction of the dense core is not superfluid/superconductor.
Our modeling of the thermal evolution of the neutron star predicts that it may still cool further, down to a temperature of $\simeq$15~eV. Measuring such a low temperature with a future observation may provide constraints on the fraction of baryons that is paired in the stellar core.
\end{abstract}



\section{Introduction}\label{sec:intro}

Neutron stars (NSs), with maximum masses up to $2-2.5~\Msun$ and radii of the order of $10-14$~km \citep[e.g.,][]{Lattimer:2012aa,Ozel:2016aa}, contain the densest stable form of matter in the observable Universe. Possibly denser matter has been produced in relativistic heavy ion collisions  \citep{Busza:2018aa}, but in a form very far from being stable while the densest form of matter in the interior of black holes is not observable.
The bulk properties of dense matter are encapsulated in the equation of state (EOS) as a relationship $P = P(\rho)$ between the pressure $P$ and the mass density $\rho$. These have been constrained by measurement of masses and radii of NSs \citep[e.g.,][]{Lattimer:2012aa,Ozel:2016aa,miller2019,riley2019}, by the observation of gravitational waves from an NS-NS merger \citep{abbott2018_NSmerger_EOS} and by combined analysis of the former with further constraints obtained from pulse-profile modeling of \nicer\ data \citep{miller2020_combi,Raaijmakers:2020aa}. These characteristics are, however, well described by zero-temperature models and miss a large part of the important properties of dense matter.

The finite temperature behavior of a system is what reveals its intimate structure and can be accessed, e.g., through transport coefficients, neutrino emission rate, or specific heat. For instance, the occurrence of fast neutrino emission by a process such as direct Urca\footnote{In direct Urca processes, neutrinos are emitted when baryons exhibit successive beta decay and electron capture reactions.} can give us information about the proton abundance in the deep NS interior \citep{Lattimer:1991aa}, or the presence of baryonic particles beyond neutrons and protons, i.e., hyperons \citep{Prakash:1992aa}, deconfined quark matter \citep{Iwamoto:1980aa}, or other exotic forms of matter \citep{Yakovlev:2001aa}. 
The specific heat is particularly elegant as it depends mostly on the properties of the fermion particle content and, unlike transport and emission processes, does not depend on the very model-dependent description of particle collisions.

Low-mass X-ray binaries (LMXBs) are binary systems in which an NS can accrete matter from a low-mass companion star, typically through a disc. Most such systems are transient where accretion outbursts, with duration ranging from weeks to decades, are separated by much longer periods of quiescence during which little or no accretion onto the NS surface occurs. In quiescence, the bare surface of the NS, not being outshined by the accretion disc, can be directly observable. 
Moreover, these NSs in LMXBs have very low surface magnetic fields that have a negligible effect on the atmosphere structure and allow for reliable measurement of the surface effective temperature \citep{Rutledge:1999aa}. 
This makes transient LMXBs unique
systems to probe the interior of NSs via surface temperature measurements \citep[e.g.,][for a recent review]{wijnands2017}.

Transient accretion leaves its marks on the thermal evolution of an NS. During an accretion outburst, the accreted matter is slowly pushed to increasing densities throughout the NS crust (i.e., the outer regions where densities are still lower than nuclear matter density) undergoing a series of non-equilibrium reactions, such as electron captures, neutron emission, and pycno-nuclear fusions \citep{Sato:1979aa,Bisnovatyi-Kogan:1979aa,Haensel:1990aa}. Together, these result in a heat release of $1.5 - 2$ MeV per accreted baryon, a process called {\it deep crustal heating} \citep{brown1998}. For long enough outbursts, this results in the NS crust being pulled out of thermal equilibrium with the stellar core (the region at supra-nuclear densities comprising about 99\% of the star's mass) \citep{rutledge2002}. 
%
%

Most of the gravitational energy of the accreted matter is liberated as heat at the surface and is immediately radiated away since the surface is colder than the underlying ocean, which is heated by the nuclear burning of hydrogen
(see e.g., \citealt{1998ASIC..515..419B}).
Nevertheless, it has been hypothesized that some fraction of this gravitational energy can be transferred deeper into the NS to densities around  $10^8 - 10^{11}\, \gcc$ \citep[][]{Inogamov:2010aa}. This could be the source for a mechanism not yet fully understood that has been dubbed as {\it shallow heating} \citep{Brown:2009aa}, although other explanations for this apparent additional heat generation have been proposed too \citep[e.g.,][]{Medin:2011aa,Steiner:2012aa}.
Moreover, in the case of such a cold star as \HETE, it cannot be excluded that part of the nuclear energy from the burning of H and He, either stable or explosive during X-ray bursts, is not wholly radiated away and is rather able to leak into the NS crust and contribute to the shallow heating.

Once an accretion outburst ceases, all the energy gained by deep and shallow heating processes is radiated away through neutrinos from the stellar core or photons from the surface. Such post-outburst cooling of NSs has been observed for several transient LMXBs by monitoring these systems for years to decades with sensitive X-ray telescopes \citep[e.g.,][]{cackett2008_mxb,degenaar2011_exo,diaztrigo2011_exo,fridriksson2011_xte,merritt2016_ks,parikh2017_maxi}. Such observations provide the opportunity to infer the structure of the NS crust \citep[e.g.,][]{wijnands2004,degenaar2011,degenaar2014_exo,ootes2019_t5x2}, including the presence of a low-conductivity pasta layer \citep[][]{Horowitz:2015aa}. In addition, such studies allow to investigate the heat capacity \citep[][]{cumming2017} and neutrino emissivity \citep[][]{brown2018} of the dense core and even have the potential to constrain the compactness of the NS \citep[e.g.,][]{Deibel:2015aa}.

\subsection{The transient NS LMXB \HETE}\label{subsec:hete}

\HETE\ was discovered as an accreting NS in 2005 \citep{vanderspek2005}. While it initially displayed X-ray pulsations, indicating that the accreted material was channeled towards the NS magnetic poles, the pulsed signal was detected only sporadically after $\sim$2 months \citep[possibly because the magnetic field was buried by the accreted material;][]{patruno2012}. Various methods to estimate the magnetic field strength of the NS suggest that it is weak, $\lesssim 10^{9}$~G \citep[][]{Mukherjee:2015aa}. 

The source continued to be seen in outburst for over a decade, until its activity ceased in late 2015 \citep{degenaar2017}.\footnote{We note that a brief episode of reduced accretion activity, lasting $\sim$2 weeks, occurred in 2007 \citep[as described in][]{degenaar2017}.} A \chan\ observation taken a few months later, in 2016 April, allowed to measure the effective surface temperature by fitting the X-ray spectral data with an NS atmosphere model. These fits showed that, despite its very long outburst, \HETE\ was strikingly cold compared to other NS-LMXBs. Thermal evolution simulations using the code \textsc{NSCool} \citep{pagereddy2013,ootes2016,Page:2016aa} revealed that during this 2016 observation the measured temperature was likely set by that of the accretion-heated crust \citep{degenaar2017}. It was noted that the source would likely cool further such that it may provide interesting constraints on the heat capacity \citep{degenaar2017,cumming2017}. Given these prospects, we observed \HETE\ again with \chan\ in 2018 June, $\approx$2.5 year after its outburst had ended. We present in this paper the results and implications of this second observation.

In \sect{sec:obs} we analyse our 2018 observation and in \sect{sec:preliminary} we present first an analytical study of its implications. Section \ref{sec:MCMC} then offers a detailed investigation through Markov-chain Monte-Carlo (MCMC) simulations, where we leave details on the numerical modeling and MCMC setup for appendices. We conclude in \sect{sec:conclusions} and provide an outlook for what future observations of \HETE\ may teach us about its interior properties.

\section{2018 OBSERVATION AND DATA ANALYSIS}
\label{sec:obs}

\subsection{Chandra data reduction and analysis}

Following our previous study of \HETE, we obtained a new \chan\ observation on 2018 June 4 (PI: Degenaar, proposal ID: 19400265). Data 
were collected between 03:04 and 19:10 UT, for a total exposure time of 55.9 ks. The ACIS-S3 chip was operated in the very faint and timed data mode. We reduced and analyzed the \chan\ data using standard tools incorporated in \textsc{ciao} (v. 4.9).\footnote{https://cxc.cfa.harvard.edu/ciao} As a first step, the data were reprocessed using \textsc{acis\_process\_events}. By extracting a 100-s binned light curve of the entire CCD, we then determined that no background flares occurred during the observation and that all data could be used in the analysis. 

To extract source counts, we used a $1$ arcsec circular region centered on the source position determined by running \textsc{wavdetect} on the 0.5--2 keV image (see Section~\ref{subsec:specsim}). The obtained coordinates, R.A. $=$19:00:08.686 and Dec. $=-$24:55:14.315 (J2000), are consistent with the optical position of the system \citep[known to sub-arcsecond precision;][]{Fox:2005aa}. A surrounding source-free annulus with an inner/outer radius of $5$/$25$ arcsec was used to estimate background counts. To facilitate spectral simulations (Section~\ref{subsec:specsim}), we created an observation-specific redistribution matrix file (rmf) and ancillary response file (arf) with \textsc{specextract}. \\

\begin{figure*}
\includegraphics[width=15cm]{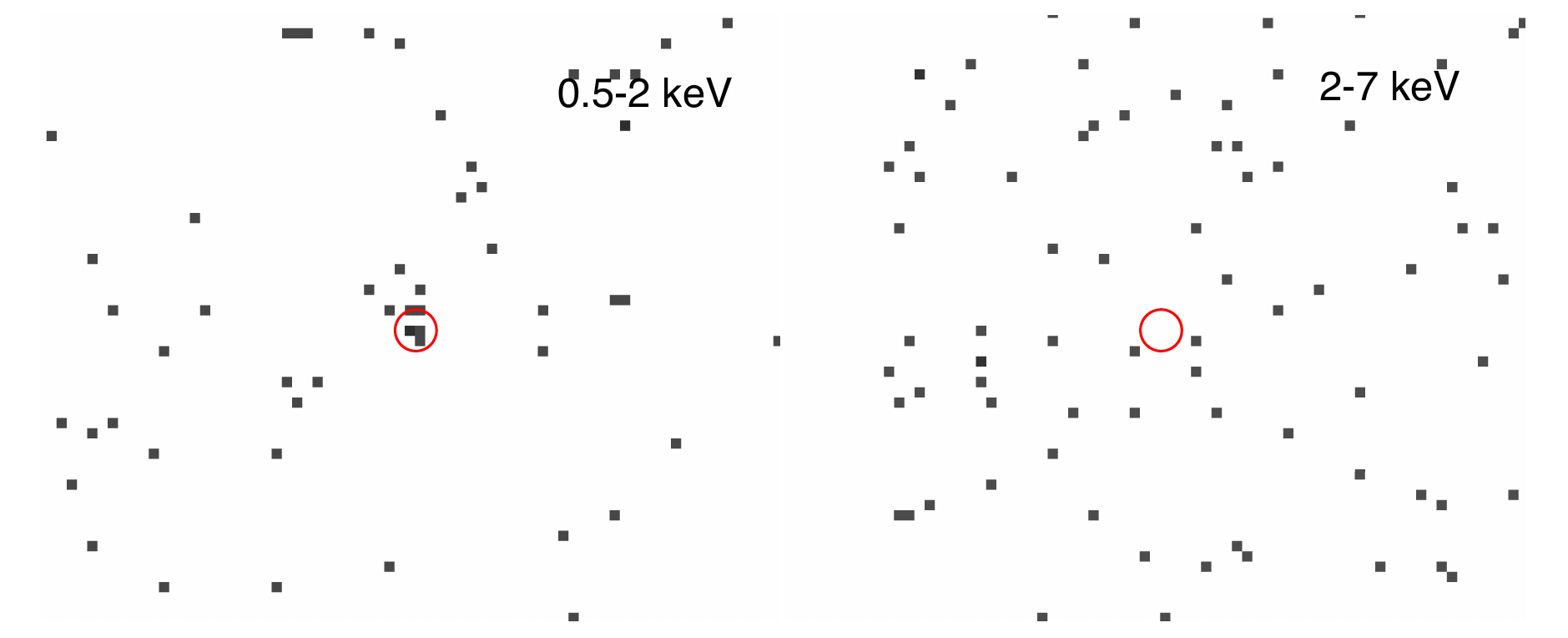}
\caption[]{\chan/ACIS images of the field around \HETE\ obtained in 2018. Shown are images in the 0.5--2 keV band (left) and in the 2--7 keV band (right). Our 1-arcsec source extraction region is indicated in red in both images.
}
 \label{fig:ds9}
 \end{figure*}   

\begin{figure}
\includegraphics[width=8.5cm]{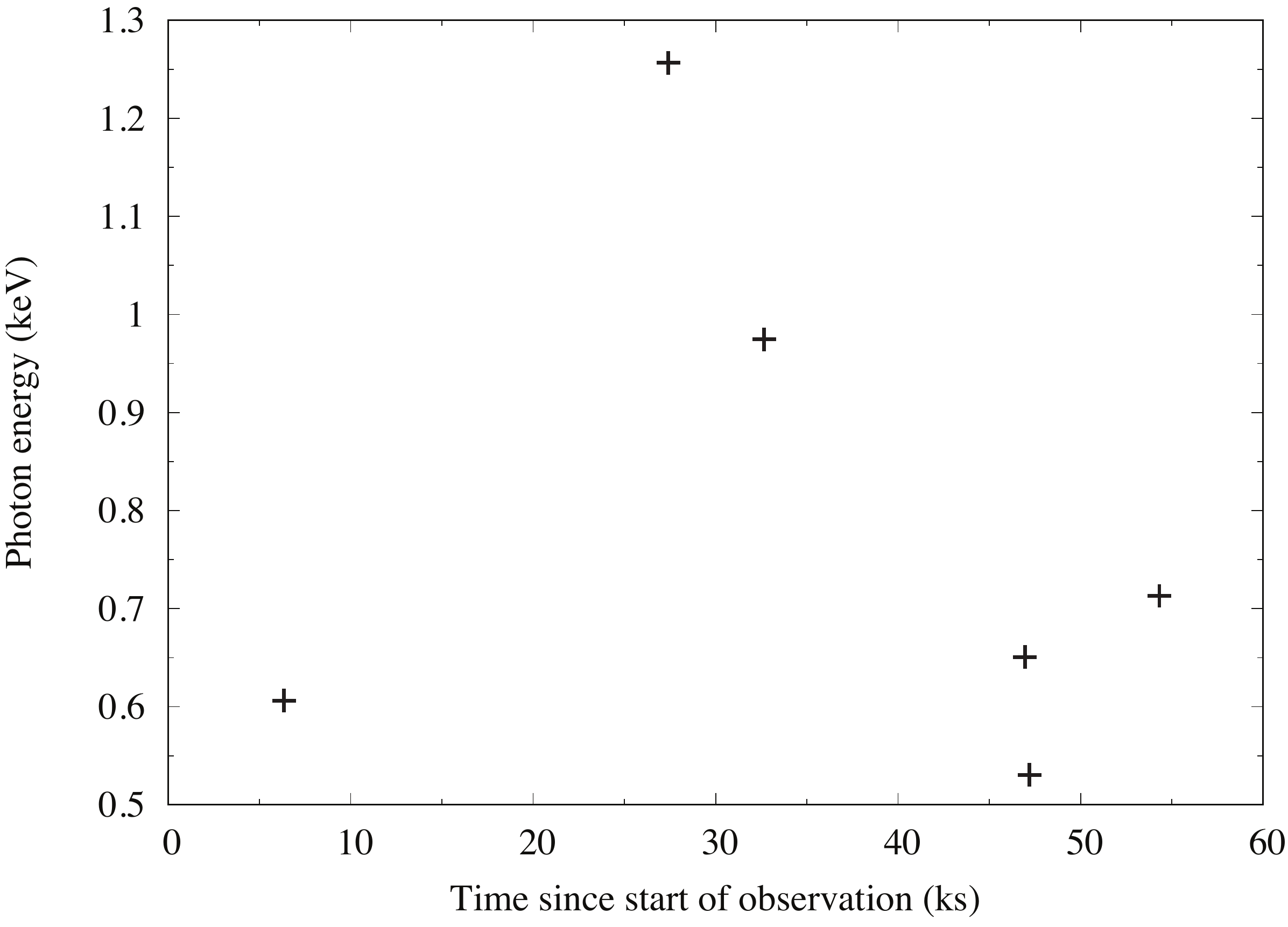}
\caption[]{
Energies of the photons detected from the position of \HETE\ versus the time along our 2018 \chan\  observation. 
}
 \label{fig:energy}
 \end{figure}   


\begin{figure}
\begin{center}
\includegraphics[width=0.47\textwidth]{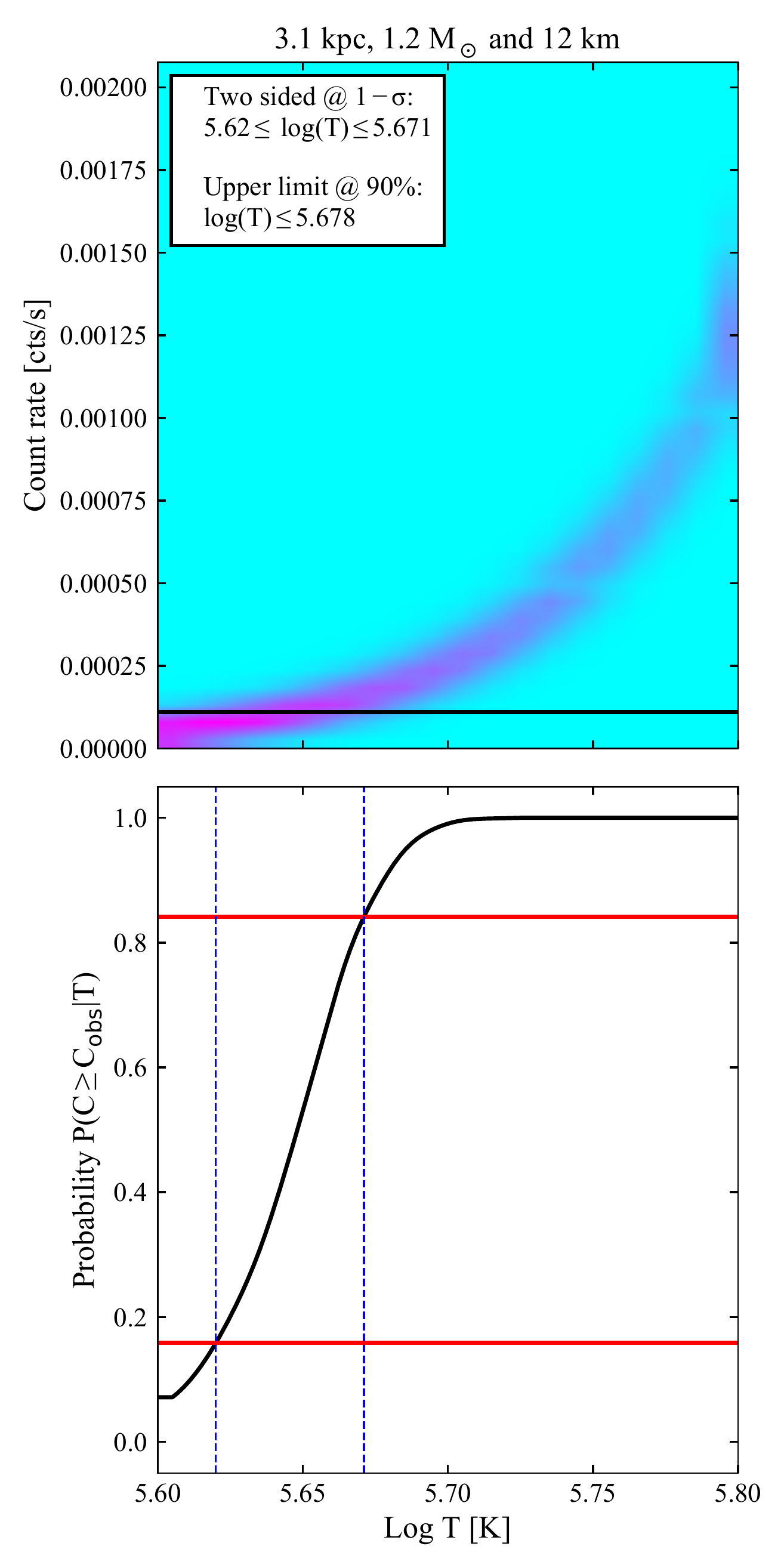}
\end{center}
\caption{Method to determine a temperature confidence limit for the observed count rate, depending on the assumed distance, mass, and radius. Shown is an example for $D=3.1$~kpc, $M=1.2~\Msun$, and $R=12$~km. Top: The interpolation of the $2\times10^4$ spectral simulations to a two-dimensional function that gives the probability for each combination of temperature and count rate. Bottom: The probability that the count rate is higher than the observed count rate, as a function of temperature. The red horizontal lines in this plot correspond to the 1-$\sigma$ upper and lower limits.
}
\label{fig:ratetotemp}
\end{figure}
 
\begin{figure*}
\begin{center}
\includegraphics[width=0.99\textwidth]{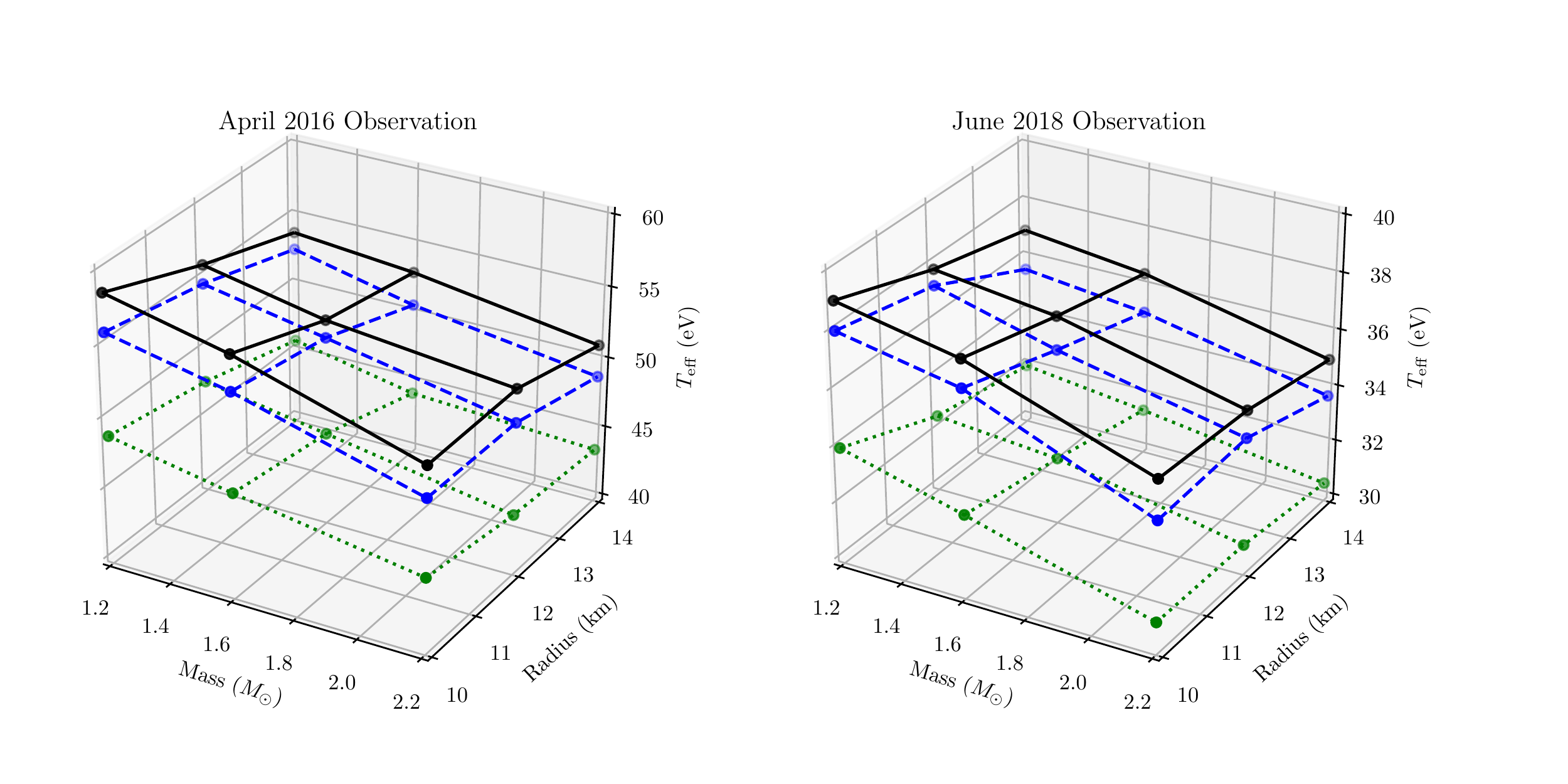}
\end{center}
\caption{A three-dimensional representation of the effective NS temperatures inferred from our 2016 (left) and 2018 (right) observations for different assumed masses and radii.
We used linear interpolation from this grid in our MCMC simulations to obtain a temperature for any mass-radius combination. 
The three grids correspond to distances $D=5.3$ kpc for continuous (black) lines, $D=4.7$ kpc for dashed (blue) ones, and $D=3.1$ kpc for dotted (green) ones.
}
\label{fig:grid}
\end{figure*}
 
\subsection{Source count rate and spectral shape}\label{subsec:specsim}

Our target is not detected by eye in the full-band image of our new 2018 observation. Extracting images in different energy bands, 0.5--2~keV and 2--7~keV, reveals that the source is detected by eye only at lower X-ray energies (see Figure~\ref{fig:ds9}). This is a first indication that the X-ray spectrum has a soft spectral shape, as expected if thermal emission from the NS is observable. Using \textsc{dmcopy}, we extracted the energies for each of the photons detected in the source region, which shows that their individual energies are between 0.5 and 1.3~keV (see Figure~\ref{fig:energy}).

We use \textsc{ciao} tool \textsc{aprates}\footnote{https://cxc.cfa.harvard.edu/ciao/ahelp/aprates.html} to determine the count rate and uncertainty for \HETE\ in the 0.5--2 keV band. Running the \textsc{ciao} tool \textsc{src$\_$psffrac}, we estimate that 92\% of the source point spread function is contained within our $1$-arcsec source extraction region and 0\% in our background region. Using these numbers as input for \textsc{aprates}, we obtain a count rate of $1.1\times10^{-4}~\cnts$ with a 1-sigma confidence region extending from $7.2\times10^{-5}~\cnts$ to $1.7\times10^{-4}~\cnts$. For comparison, using \textsc{aprates} for our 2016 observation, we obtain a count rate of $1.1\times10^{-3}~\cnts$ with a 1-sigma confidence region extending from $9.4\times10^{-4}~\cnts$ to $1.2\times10^{-3}~\cnts$ (0.5--2 keV). This shows that the ACIS-S3 count rate of our target decreased by a factor of $\approx$10, over the $\approx$2.1 yr that separates our two observations. 
Although the soft-energy response of the ACIS-S is known to change over time \citep[e.g.,][]{Plucinsky:2016}, these are variations at the percent level \citep[][]{posselt2018}. The large change in count rate that we observe for \HETE\ can thus not be explained by instrumental effects and hence shows that \HETE\ faded between our two observations.

\subsection{Temperature determination}
\label{sec:temp}

Our 2018 observation does not collect sufficient photons for detailed spectral analysis. To infer the temperature of the NS, we therefore compared the measured count rate to spectral simulations, performed using \textsc{XSpec} (v. 12.9).\footnote{https://heasarc.gsfc.nasa.gov/docs/xanadu/xspec/index.html} To this end, we first motivate our choice of spectral model for the simulations.

Our earlier 2016 \chan\ observation allowed for spectral analysis and this revealed that, at that time, the quiescent spectrum of \HETE\ had a soft thermal shape with no evidence for a hard emission tail \citep[][]{degenaar2017}. Since our imaging analysis of the 2018 \chan\ observation also suggests a soft X-ray spectrum (see Section~\ref{subsec:specsim}), it is reasonable to assume that it maintained the shape of an NS atmosphere model. The source shows no evidence of a high inclination; both X-ray and optical studies performed during its outburst point to a low inclination \citep[e.g.,][]{Elebert:2008,Papitto:2013}, and the 2018 quiescence spectrum did not reveal any heavy absorption \citep[][]{degenaar2017}. It is therefore further reasonable to assume that the absorption along the line of sight, the hydrogen column density, $N_{\mathrm{H}}$, did not change between our two observations. 

Based on the results of our 2016 \chan\ observation, we model the source spectrum with an NS atmosphere model that is subject to interstellar absorption. As in our previous analysis, we choose the frequently used model \textsc{nsatmos} \citep{heinke2006}. We further used the \textsc{tbabs} model for the interstellar extinction, adopting \textsc{vern} cross-sections \citep{verner1996} and \textsc{wilm} abundances \citep{wilms2000}. To explore our parameter space, we performed these spectral fits for three different values of the distance ($D=3.1$, 4.7, and 5.3~kpc),\footnote{These three values correspond the minimum, most likely, and maximum source distance based on the analysis of thermonuclear bursts \citep[][]{Galloway:2008aa}. We note that the distance estimate cannot be improved with \gaia\ due to the optical faintness of the source; the parallax listed in \gaia\ Data Release 2 has a large error of $0.325 \pm 0.288''$.} mass ($M=1.2$, 1.6, and 2.2~M$_{\odot}$), and radius ($R=10$, 12, and 14~km)
\footnote{Unless otherwise specified by radius, $R$, we mean the areal coordinate such that the surface area is $4\pi R^2$,
to be distinguished from the ``radius at infinity'', $R^\infty \equiv e^{-\phi(R)}R$ (see Section 3 for the definition of the redshift notation used here).}.

For each triplet of $D$, $M$ and $R$, we used the \textsc{XSpec} routine \textsc{fakeit}, adopting the observation-specific rmf and arf files, to simulate spectra for different temperatures. Based on the measurement of our previous \chan\ observation, we choose a range running from $\log T = 5.6$ to 5.9, with a step size of $\log T = 0.01$. For each temperature, we simulated $2\times 10^4$ individual spectra and determined for each single spectrum the resulting ACIS-S3 count rate in the 0.5--2 keV energy band using the \textsc{XSpec} command \textsc{show rates}.

We converted the obtained distribution of simulated count rates for every chosen temperature into a temperature constraint, for different combinations of $D$, $M$, and $R$. For this purpose, we used a custom \textsc{python} script to compute a two-dimensional (2D) cubic spline interpolation of the simulated count rate distribution as a function of temperature. Using the 2D interpolation, we could calculate the probability of observing a count rate equal to or higher than the observed count rate for any temperature. We then used a Nelder-Mead algorithm to minimize the difference between that probability and the two-sided 1$\sigma$ lower and upper limits (i.e. $\approx (1-0.682689)/2$ and $\approx (1+0.682689)/2$) with respect to temperature. We repeated this procedure for all combinations of distance, mass, and radius. An example of this analysis is shown in Figure~\ref{fig:ratetotemp}. The grid of temperatures for different masses and radii that we obtained in this way is shown in Figure~\ref{fig:grid} (right) and in the left-hand panels, we include the analogous results for the 2016 data obtained, in that case, from spectral fits. 
For 2018, it ranges from $\approx39$~eV for the combination of the largest distance and lowest mass and radius to $\approx30$~eV for the combination of our lowest distance and maximum mass and radius.

\section{An analytical study of \HETE}
\label{sec:preliminary}

We present in this section analytical estimates of the properties of the dense stellar core based on our new temperature measurement of \HETE\ and its observed outburst properties. 
We consider a spherical star, with the standard metric obtained by solving the Tolman-Oppenheimer-Volkoff equations (see e.g.,
\citealt{ShapiroTeukolsky}), which we assume to be in thermal equilibrium, i.e., with a uniform internal redshifted temperature 
$\widetilde{T} = e^{\phi(r)} T(r)$, where
$e^{\phi(r)}$ is the redshift factor inside the NS and $T(r)$ the local temperature.
Similarly we denote redshifted energies as $\widetilde{E}$ and doubly-redshifted\footnote{Luminosity is energy per time: energy is redshifted and time is blue-shifted so "per time" is redshifted from which a double redshift results.}
luminosities as $\widetilde{L}$
so that general relativistic effects are automatically taken care of with this ``tilde'' notation.

\subsection{Heat capacity and superfluid properties}
\label{sec:CV}

\begin{figure}[t]
\begin{center}
\includegraphics[width=0.47\textwidth]{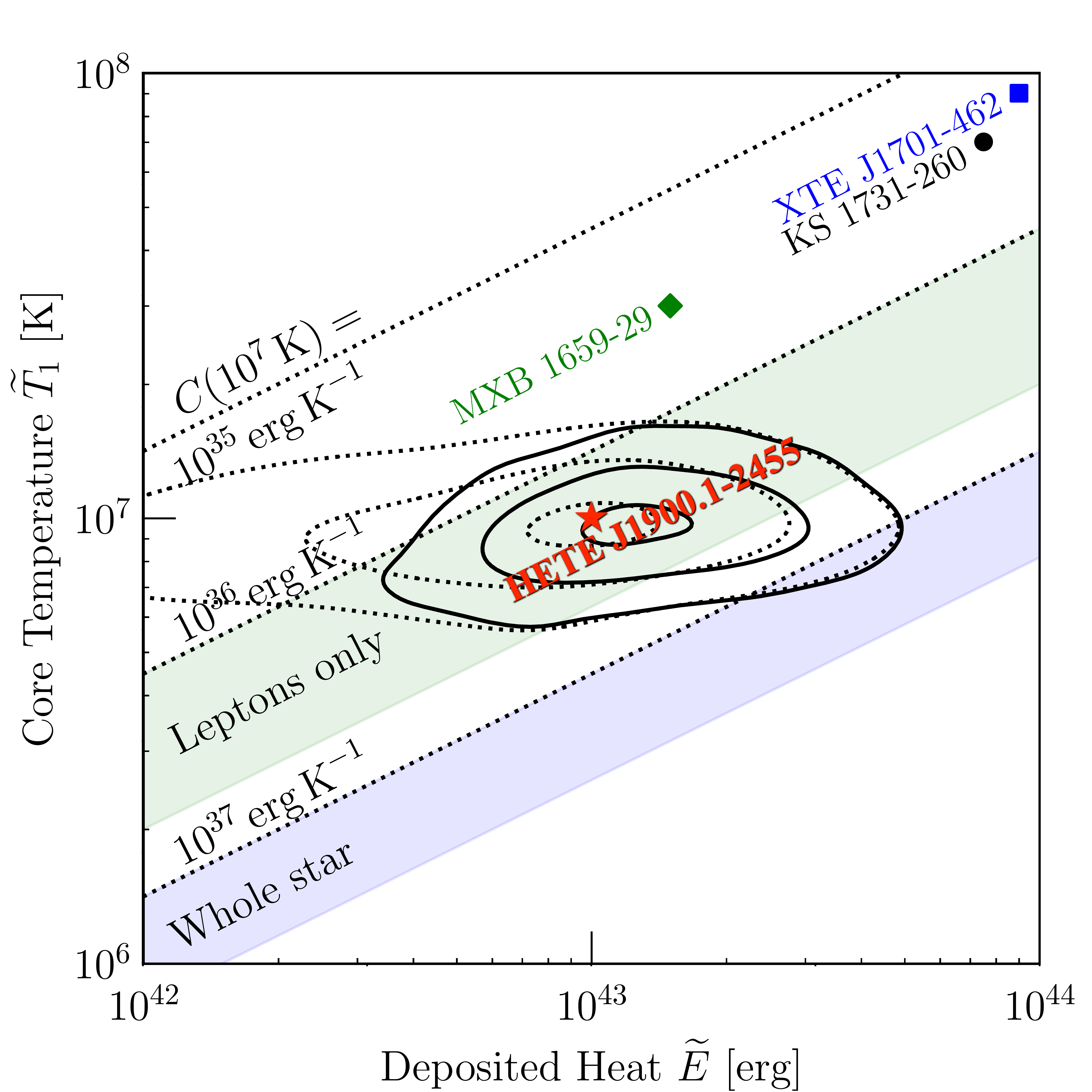}
\end{center}
\caption[]{
Location in the $\widetilde{E}$ - $\widetilde{T}_1$ plane of the estimated values of generated heat, $\widetilde{E}_\mathrm{h}$, and core temperature, $\widetilde{T}_1$, of the NS in \HETE\ at the time of our 2018 \chan\ observation, compared to similar results from three other systems \citep[][]{cumming2017}. 
The three dotted lines show the lower limits, from \eq{Eq:lowlim}, on the total stellar heat capacity $C_7 \ge 2\widetilde{E}_\mathrm{h}/\widetilde{T}_1 \cdot (10^7 \, \mathrm{K}/\widetilde{T}_1)$ evaluated at an internal redshifted temperature of $\widetilde{T} = 10^7$ K. 
Colored in green and blue are the expected ranges of $C_7$ originating from only the leptons and from the whole star, i.e., leptons+unpaired nucleons, at the same temperature, see Equations (\ref{Eq:Clep}) and (\ref{Eq:Cnucl}).
The three black contours show, from the inner to the outer one, the $1\sigma$,  $2\sigma$, and $3\sigma$ confidence levels
from the position of the cooling models of our MCMC run B delimiting the range of inferred values of deposited energy $\widetilde{E} = \widetilde{E}_\mathrm{h}$ and $\widetilde{T}_1$ for \HETE\ at the time of our 2018 observation. The three dotted contours show the same but when considering neutrino losses, i.e., with $\widetilde{E} =\widetilde{E}_\mathrm{eff} \equiv \widetilde{E}_\mathrm{h}-\widetilde{E}_\nu$.
}
\label{fig:ALL1}
\end{figure}

It was noted by \citet{pagereddy2013} that, in the case of a strong enough or a long enough accretion outburst,
not only the crust but also the stellar core may see its temperature increase.
This opened the possibility to use NS-LMXBs as a probe of the NS heat capacity, $C$, as described in detail in \citet{cumming2017}.
The heat capacity of an NS is strongly dominated by its core and is predominantly provided by excited states of its fermionic components (i.e., particles with half-integer spin), with bosonic excitations (i.e., integer spin particles) making only a negligible contribution. 

Fermionic components naturally present in an NS core are neutrons and protons, while electrons and muons (i.e., leptons) are needed to guarantee charge neutrality. Detailed calculations \citep{Ofengeim:2017aa,cumming2017} show that for a wide range of EOSs and stellar masses lower than $\approx 2.2~M_\odot$, 
the lepton contribution to the heat capacity is in the range of
\be
C^\mathrm{lep} \approx (1-5) \times 10^{36} \, \widetilde{T}_7 \; \mathrm{erg \, K}^{-1} \; ,
\label{Eq:Clep}
\ee
while the contribution of unpaired (i.e., non-superfluid; see below) nucleons is in the range 
\be
C^\mathrm{nucl} \approx (1-3) \times 10^{37} \, \widetilde{T}_7 \; \mathrm{erg \, K}^{-1} \; .
\label{Eq:Cnucl}
\ee
In both cases the heat capacity is assumed to be proportional to $\widetilde{T}$, as is the case for strongly degenerate fermions \citep{Baym:2004aa}, and $\widetilde{T}_7$ denotes $\widetilde{T}$ in units of $10^7$~K.

At high densities, i.e., high Fermi energies, matter is expected to take more exotic forms; hyperons (i.e., baryonic particles that contain a strange quark) and deconfined quarks may appear.
However, in both cases, the threshold appearance density remains unknown (see e.g., \citealt{Page:2006aa} for a review).
Degenerate hadrons (i.e., all particles made out of quarks), having strong attractive interactions are, however, likely to undergo a Cooper pair instability \citep{Cooper:1956aa} resulting in superfluidity/superconductivity \citep{Bardeen:1957aa}. 
Such pairing is expected to occur in the dense matter of the NS interior \citep{Page:2014aa,Gezerlis:2014aa}.
The formation of a condensate of Cooper pairs lowers the energy of the system, i.e., this becomes the ground state while single particles present an excited state separated by a certain energy gap, the energy needed to break a pair.
The appearance of an energy gap in the single-particle excitation spectrum results in a strong (often exponential) suppression of the specific heat \citep{Levenfish:1994aa}.
A measurement of, or constraint on, an NS's heat capacity $C$ can thus provide a handle on the extent of pairing in its core:
the higher $C$ is, the smaller is the extent of pairing.

To see how observations of NS-LMXBs can be used to put constraints on the core heat capacity, let us consider the thermal evolution of the star from an initial uniform redshifted temperature $\widetilde{T}_0$, through an accretion outburst during which a total amount of heat $\widetilde{E}_\mathrm{h}$ was injected into the star, and follow it till the thermal disequilibrium between the crust and the core has relaxed so that the star is finally at a new uniform, redshifted, temperature $\widetilde{T}_1$. We can write its heat capacity from degenerate fermions as $C(\widetilde{T}) = \widetilde{D} \cdot \widetilde{T}$,
where $\widetilde{D}$ is a $T$-independent quantity, and the thermal energy as $\widetilde{E}_\mathrm{th}(\widetilde{T}) = \frac{1}{2} \widetilde{D} \cdot \widetilde{T}^2$. If we had measurements, or at least reliable estimates, of $\widetilde{T}_0$, $\widetilde{T}_1$, and $\widetilde{E}_\mathrm{h}$, then we could simply obtain $\widetilde{D}$ from
\be
\Delta \widetilde{E}_\mathrm{th} = \widetilde{E}_\mathrm{h} = \frac{1}{2} \widetilde{D} (\widetilde{T}_1^2 - \widetilde{T}_0^2) \; .
\ee
Unfortunately, there are no cases where strong heating of an NS core occurred and a final isothermal state $\widetilde{T}_1$ has been reached, for which we have a known initial $\widetilde{T}_0$. So, to date, the best we can obtain is a lower limit
\be
C(\widetilde{T}_1) = \widetilde{D} \widetilde{T}_1 \ge \frac{2 \widetilde{E}_\mathrm{h}}{\widetilde{T}_1}
\label{Eq:lowlim}
\ee
obtained by considering the extreme case where $\widetilde{T}_0 \ll \widetilde{T}_1$.

\citet{cumming2017} used measurements of the core temperatures and accretion energetics of three NS-LMXBs with long outbursts and cold cores (\ks, \mxb, and \xte), to explore the resulting constraints on the NS heat capacity (see \fig{fig:ALL1}). 
This methodology was also applied to \HETE\ and its $T_\mathrm{eff}^\infty \approx 55$ eV inferred from the 2016 observation \citep{degenaar2017}. 
For all four sources, it was found that the lower limits on $C$ were still a factor of a few below the heat capacity provided by leptons only (Equation~\ref{Eq:Clep}). This implies that those data do not provide constraints on the level of baryon pairing.

We can now consider the consequence of the much lower $T_\mathrm{eff}^\infty$ of \HETE\ from the 2018 observation. We adopt $T_\mathrm{eff}^\infty \approx 35$~eV, as is appropriate for the parameter combination $D=4.7$~kpc, $M=1.6~\Msun$, and $R=12$~km (i.e., the middle of the full range we consider; see Figure~\ref{fig:grid}). 
From the right-hand panel of figure~3 in \cite{degenaar2017}, which shows the mapping between the observed surface temperature and that of the NS interior, one then immediately deduces\footnote{As can be seen in figure~3 of \cite{degenaar2017}, for the very cold NS in \HETE\ the relation between these temperatures is independent of the unknown amount of light elements in the NS envelope as long as it is above $\sim 10^6$ g cm$^{-2}$. These light elements present in quiescence were previously deposited during the accretion phase and, at a rate of 1\% of the Eddington rate 
with $\dot{m}_\mathrm{Edd}\approx 10^5$ g cm$^{-2}$, it took less than an hour to accrete this minimum of $\sim 10^6$ g cm$^{-2}$.} an internal temperature $\widetilde{T}_1 \approx 10^7$~K. With an average mass accretion rate of $\sim 2.3\times 10^{16} \mdotg$ \citep{degenaar2017} and assuming an injected energy $Q \sim 2$ MeV per accreted baryon, the total heat deposited into \HETE's interior during its 10 yr long outburst is $\widetilde{E}_\mathrm{h} \sim 10^{43}$ ergs.
Employing these values, one obtains an approximate lower limit from \eq{Eq:lowlim} of
\begin{equation}
C_7 \equiv C(\widetilde{T}=10^7 \text{K}) \gtrsim 2 \times 10^{36} \, \text{erg K}^{-1} \, .
\label{Eq:CV_approx}
\end{equation}
This rough estimate is already in the range of theoretically expected values of $C^\mathrm{lep}$, see \eq{Eq:Clep},
meaning that we have in \HETE\ a system that may allow us to test some basic tenets of physics. We show in \fig{fig:ALL1} the location of \HETE's estimated values of $\widetilde{E}_\mathrm{h}$ and $\widetilde{T}_1$ in an $E-T$ plane and compare it with the three other NS-LMXBs used by \citet{cumming2017}. Our new observation appears to provide us with a one order-of-magnitude improvement on the lower limit of an NS's heat capacity.

\subsection{Outburst recurrence time and core neutrino emission}
\label{sec:nu}

On time-scales much longer than the outburst and crust relaxation time, it is likely that the star will maintain a balance between the average heating luminosity, $\langle \widetilde{L}_\mathrm{h} \rangle$, and the average cooling luminosity $\langle \widetilde{L}_* \rangle$, from photons and/or neutrinos \citep{brown1998,colpi2001}. Under the simplifying assumption that accretion outbursts occur regularly with a recurrence time $\tau_\mathrm{rec}$, one can consider for practical estimates that $\langle \widetilde{L}_\mathrm{h} \rangle \approx \widetilde{E}_\mathrm{h}/\tau_\mathrm{rec}$.

We have no direct information on the recurrence time $\tau_\mathrm{rec}$ of \HETE, but if one observed outburst of length $\tau_\mathrm{ob} \simeq 10$ yr in 50 yr of X-ray astronomy is indicative, then $\tau_\mathrm{rec} >$ 50 yr.
However, it is possible that a previous outburst in the past century has been missed. Perhaps a more conservative lower limit would be $\tau_\mathrm{rec} >$ 20 yr, which is roughly since when all-sky X-ray monitoring has been happening continuously (with various instruments). 
In many LMXBs, $\tau_\mathrm{rec}$ is estimated to be of the order of 10--100 times the outburst time $\tau_\mathrm{ob}$ \citep[e.g.,][]{yan2015}. Indeed, the NS systems Aquila X-1 and \sax, which have the lowest \citep{Ootes:2018aa} and highest \citep{Heinke:2007aa} inferred neutrino efficiencies, respectively, and have exhibited several outbursts over the past decades, both have $\tau_\mathrm{rec} \approx 10 \times \tau_\mathrm{ob}$. As another example, the crust-cooling source \mxb\ has exhibited three outbursts, with a summed duration of $\simeq$ 6.5 yr, since its discovery in 1976 and thus appears to have $\tau_\mathrm{rec} \approx 7 \times \tau_\mathrm{ob}$ \citep{Parikh:2019aa}. If \HETE\ has a similar duty cycle of $\sim$10\%, we would thus expect $\tau_\mathrm{rec} \sim 100$~yr. Bearing this in mind, we can now consider the thermal state of the NS in view of its outburst history.

Assuming that all thermal X-ray emission detected from \HETE\ in quiescence ($\widetilde{L}_\gamma \approx 3 \times 10^{31}$ erg s$^{-1}$) is re-radiation of heat that was deposited there during accretion episodes ($\widetilde{E}_\mathrm{h} \sim 10^{43}$~ergs; see Section~\ref{sec:CV}) would suggest that it exhibits an outburst every $\tau_\mathrm{rec} \approx \widetilde{E}_\mathrm{h}/\widetilde{L}_\gamma \approx 10^4$ yr. This is three orders of magnitude larger than the outburst time $\tau_\mathrm{ob} = 10$ yr. For \HETE\ to have a similar outburst recurrence time as other (NS) LMXBs would thus require that a large fraction of the energy deposited during accretion episodes is radiated away by other means, i.e., through neutrinos emitted from its dense core. 
At internal temperatures $T \approx 10^7$~K, the neutrino luminosity from the standard modified-Urca processes is even lower than the photon luminosity and negligible \citep{Page:2006ab}. So, unless $\tau_\mathrm{rec}$ is unusually long (as estimated above), \HETE\ must be undergoing more efficient neutrino core cooling. 
To gauge this, let us consider the case of the direct Urca (DU) process \citep{Lattimer:1991aa}, which results in $\widetilde{L}_\nu^\mathrm{DU} \approx x_\mathrm{DU} 10^{35} \widetilde{T}_7^6$ erg s$^{-1}$, where $x_\mathrm{DU}$ is the volume fraction of the core where DU is acting.
In that case, $\tau_\mathrm{rec} \approx 3 x_\mathrm{DU}^{-1} \,\widetilde{T}_7^{-6}$ yr
and hence a more common duty cycle, resulting in a recurrence time of a few decades/centuries, is readily possible.

Based on these considerations, we conclude that \HETE\ is a strong candidate to exhibit enhanced neutrino cooling. This is exciting because the rate of neutrino cooling depends on what particles are present in the stellar core and whether or not these are paired in a superfluid  \citep[e.g.,][]{Page:1990aa,Page:1992aa,yakovlev2004,Page:2006ab,Page:2006aa}. Establishing that rapid neutrino cooling is occurring in an NS core can probe what fraction of it is not superfluid/superconducting \citep[e.g.,][]{ho2015}, and what the relative fraction of protons is \citep[which gives information about the symmetry energy relevant for nuclear physics; e.g.,][]{horowitz2014}, or the presence of some exotic form of matter as hyperons or quarks, 
all of which provide valuable insight into the behavior of neutron-rich, ultra-dense matter \citep[see also][]{brown2018}. 

While exciting in itself, the possibility of the occurrence of fast neutrino emission in the inner core of \HETE\ weakens, however, our lower limit on $C$ in \eq{Eq:lowlim} and (\ref{Eq:CV_approx}) as the bound is not anymore simply from $\widetilde{E}_\mathrm{h}$ but rather from an {\it effective} heating $\widetilde{E}_\mathrm{eff} = \widetilde{E}_\mathrm{h} -\widetilde{E}_\nu$, where $\widetilde{E}_\nu$ is the energy lost to neutrinos during the process. This bound could be much lower, and much less interesting, than the one naively obtained from just $\widetilde{E}_\mathrm{h}$.

\section{A Markov-Chain Monte Carlo Exploration of \HETE}
\label{sec:MCMC}

To assess more quantitatively the possibility of a realistic constraint on the heat capacity and the occurrence of fast neutrino emission, we need to perform thermal evolution simulations. These need to model both the heating of the NS during outburst and its cooling during the subsequent phase of quiescence. For the purpose of the present work, we performed new thermal evolution simulations that significantly expand the models previously described for \HETE\ in \cite{degenaar2017}.

\begin{figure}
\begin{center}
\includegraphics[width=0.99\columnwidth]{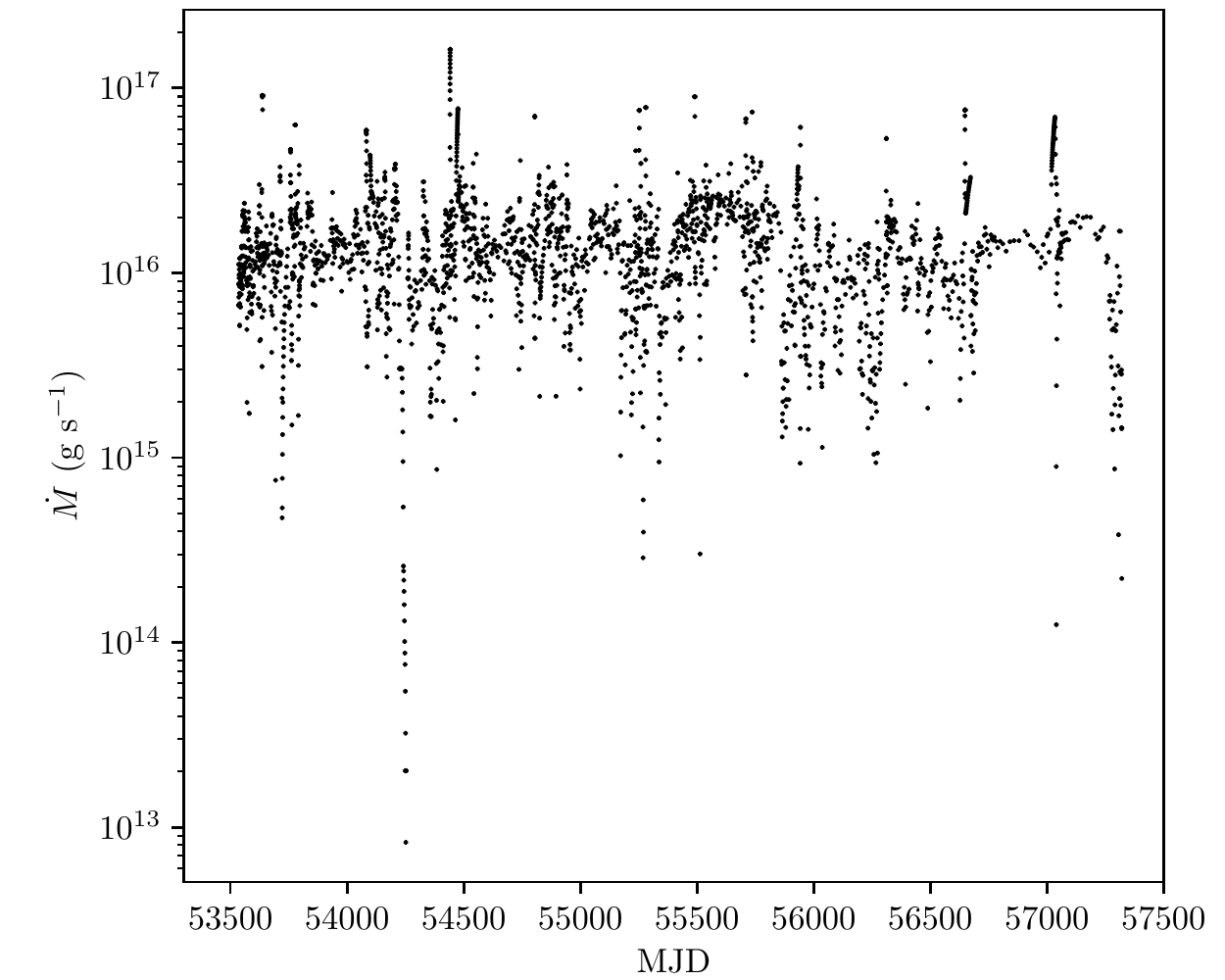}
\end{center}
\caption{Mass-accretion rate during \HETE's outburst as calculated from the all-sky X-ray monitors \rxte/ASM, \swift/BAT, and \maxi\ (see Section~\ref{sec:MCMC}), which was used as input for our thermal evolution calculations. The range of this graph corresponds to the onset (in 2005) and end (2016) of the accretion outburst of \HETE.
}
\label{fig:Mdot}
\end{figure}

The data that we use in our thermal evolution calculations are the same as described in \cite{degenaar2017}, with the addition of our new 2018 \chan\ data point. In brief, to follow the thermal evolution of the NS temperature during the accretion outburst, we make use of publicly available X-ray light curves from the \rxte/All-Sky Monitor (ASM; 2–-10 keV)\footnote{http://xte.mit.edu/ASM$\_$lc.html}, the Monitor of All-sky X-ray Image (\maxi)\footnote{http://maxi.riken.jp/top/index.html} \citep[2--20 keV;][]{Matsuoka:2009}, and the Swift/Burst Alert Telescope (BAT) transient monitor\footnote{https://swift.gsfc.nasa.gov/results/transients/}  \citep[15–-50 keV;][]{Krimm:2013}. These light curves are used to estimate the (evolution of the) mass-accretion rate during the outburst, to which the heating of the NS is presumed to be proportional. As described in \cite{degenaar2017}, the instrument count rates were first converted to Crab units and then to bolometric fluxes by assuming a bolometric correction factor of $c_{\mathrm{bol}} = 2$ \citep[][]{Galloway2008} and an accretion efficiency of $\eta = 0.2$. 
The result of converting these public X-ray monitoring light curves into a mass-accretion rate is shown in Figure~\ref{fig:Mdot}.

To constrain the thermal evolution of the NS in the subsequent quiescent phase, we make use of the fact that the source was not detected in \swift\ observations performed in 2016 March--April, i.e., in the run-up to our first \chan. In addition, we include a temperature upper limit obtained from a \swift\ observation in 2007, when the source appeared to experience a very brief period of quiescence \citep[as described in][see also Figure~\ref{fig:Mdot}]{degenaar2017}. Finally, we use the temperature constraints of our two \chan\ observations performed in 2016 and 2018.

\subsection{Modelling setup}

\begin{figure}
\begin{center}
\includegraphics[width=0.99\columnwidth]{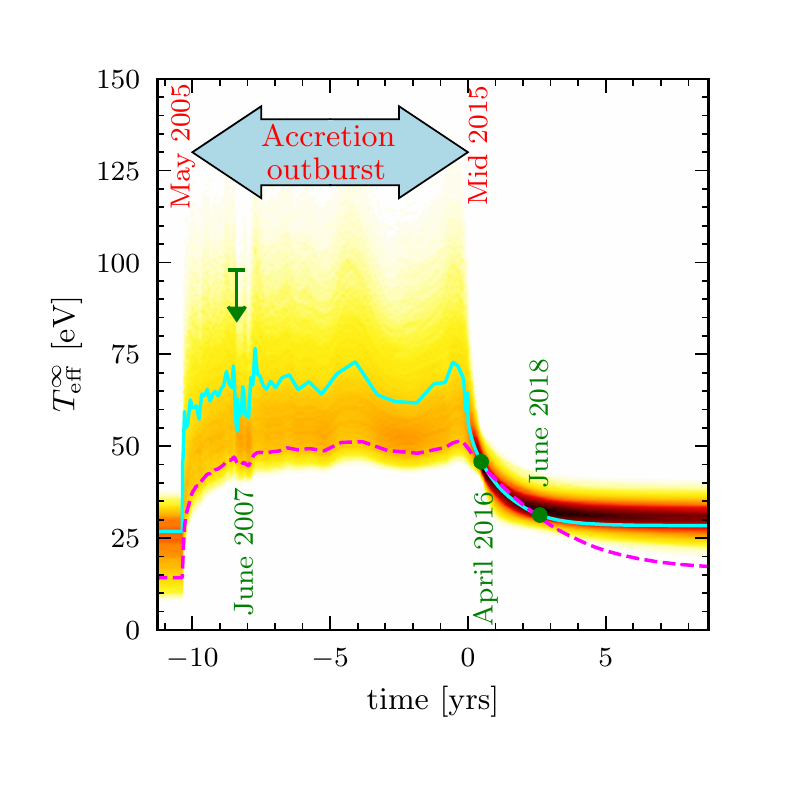}
\end{center}
\vspace{-1cm}
\caption{Generic sample of our thermal evolution models for \HETE\ during outburst ($-10.3 \, \text{yr}<t<0$) and in quiescence ($t>0$).
This plot contains about 22000 models extracted  from our MCMC run A. The yellow to red to black colors show the density of the curves. The continuous cyan curve is an example of one of our best fits and the dashed magenta curve is one of the coldest possible models. The green upper limit and the two green filled circles show the three observational constraints.
}
\label{fig:Sample}
\end{figure}

Based on our previous experiences in modelling the thermal evolution of NSs in LMXBs \citep[e.g.,][]{Parikh:2017aa,Ootes:2019aa,Degenaar:2019aa}, we performed extensive Markov chain Monte Carlo (MCMC) runs with our driver \textsc{MXMX} (see e.g., \citealt{Lin:2018aa,Ootes:2019aa})
and our NS cooling code \textsc{NSCool}. In these simulations, we take into account various observational and theoretical uncertainties in both the data analysis and the model definition. We generated more than 10 million heating-cooling curves of the accreting NS and display in \fig{fig:Sample} a generic sample of these. 

In the MCMC, we employ a set of 18 parameters. The first 16 parameters are: (1) distance $D$, (2) mass $M$, and (3) radius $R$ of the NS, (4) initial redshifted (uniform) internal temperature $\widetilde{T}_0$, 
(5-9) impurity parameter $Q_\mathrm{imp}^{(i=1 \dots 5)}$ in five crustal density ranges, (10) column density of light elements in the envelope, $y_\mathrm{L}$,
(11) strength of the shallow heating $Q_\mathrm{sh}$ and (12,13) density range over which it is acting, 
$\rho_\mathrm{sh}^\mathrm{min} - \rho_\mathrm{sh}^\mathrm{max}$,
(14,15) density range of neutron superfluidity, $\rho_\mathrm{SF}^\mathrm{min} - \rho_\mathrm{SF}^\mathrm{max}$, and
(16) fraction $a_\mathrm{entr}$ of entrained inner crust superfluid neutrons.  We provide details on all model parameters, including a motivation for the ranges used for each, in \app{app:setupMCMC}.

In addition to the 16 parameters listed above, the two most important parameters in our model are those that describe the fast neutrino emission and the core heat capacity.
The direct-Urca process between nucleons is not the only possible fast neutrino emission process.
Of comparable efficiency are the same direct Urcas involving hyperons \citep{Prakash:1992aa} or deconfined quarks \citep{Iwamoto:1980aa}.
Similar processes in the presence of charged meson condensates also have a $\propto T^6$ temperature dependence but with an efficiency
reduced by one to two orders of magnitude \citep{Maxwell:1977aa,Brown:1988aa,Tatsumi:1988aa}.
We thus apply a generic fast neutrino luminosity as
\begin{equation}
\widetilde{L}_\nu^\mathrm{Fast} = X_\mathrm{Fast} \times 10^{35} \; \widetilde{T}_7^6 \; \mathrm{erg \, s^{-1}}
\label{Eq:L_Fast1}
\end{equation}
considering values of $\mathrm{Log}_{10} X_\mathrm{Fast}$ from $-6$ to $0$.

For the core heat capacity, we write it as 
$C^\mathrm{core} = C^\mathrm{core}_7 \times (\widetilde{T}/10^7 \, \mathrm{K})$,
assuming a linear dependence on $T$ as applies to degenerate fermions \citep{Baym:2004aa}, with
\begin{equation}
C^\mathrm{core}_7 \equiv C^\mathrm{core}(\widetilde{T}=10^7 \, \mathrm{K}) = 10^\Gamma \; \mathrm{erg \, K^{-1}}
\label{Eq:C_core}
\end{equation}
and values of the exponent $\Gamma$ between $35$ up to $37.5$.
Within baryonic models the minimal expected value of $\Gamma$ is $\approx 36$ and corresponds to the case
where the whole baryon contribution is strongly suppressed by pairing and the heat capacity is fully set by the leptons' contribution, see \Eq{Eq:Clep}.
However, in the presence of an extended region of deconfined quark matter in the color-flavor locked (CFL) phase
that contains no leptons and has vanishingly small specific heat \citep{Alford:1999aa},
the core heat capacity may be smaller than the lepton contribution: we allow for this possibility
by considering values of $\Gamma$ down to 35.
Values of $\Gamma$ below 35 had, moreover, already been excluded by the study of \citet{cumming2017}.
The upper value of $\Gamma = 37.5$ corresponds to the theoretically expected value for the most massive NSs when the whole baryon contribution is accounted for (i.e., no pairing suppression is occurring). We refer to \citet{Ofengeim:2017aa} and \citet{cumming2017} for details.

We point out here that we probed two different schemes, A and B, for the electron thermal conductivity, $K_e$, in the crust. This is the main parameter that determines how the heat that is generated locally in the crust will spread over the rest of the NS, and it is controlled by the impurity parameter $Q_\mathrm{imp}$. At very low temperature, $K_e \propto Q_\mathrm{imp}^{-1}$, and we consider two possibilities for our prior on $Q_\mathrm{imp}$: either linear in the range $0$--$100$ (MCMC runs A) or logarithmic with $\mathrm{Log}_{10} Q_\mathrm{imp}$ covering (linearly) the range of $-2$ to $+2$ (MCMC runs B). 
Overall, our models are set up conservatively, covering a very large number and range of physical and astrophysical uncertainties.

For each model we estimate its recurrence time as the age $t_\mathrm{rec}$, 
counting from the beginning of the accretion outburst, at which the star has reached again an isothermal state with
temperature $\widetilde{T}$ equal to its initial $\widetilde{T}_0$ so that it could restart a new identical
accretion cycle.
Since we argued in \S~\ref{sec:nu} that the recurrence time of \HETE\ is unlikely to be shorter than 20 yr 
or longer than $10^4$ yr we discard models for which  $t_\mathrm{rec}$ is outside this range.

\begin{figure*}
\begin{center}
\includegraphics[width=0.99\textwidth]{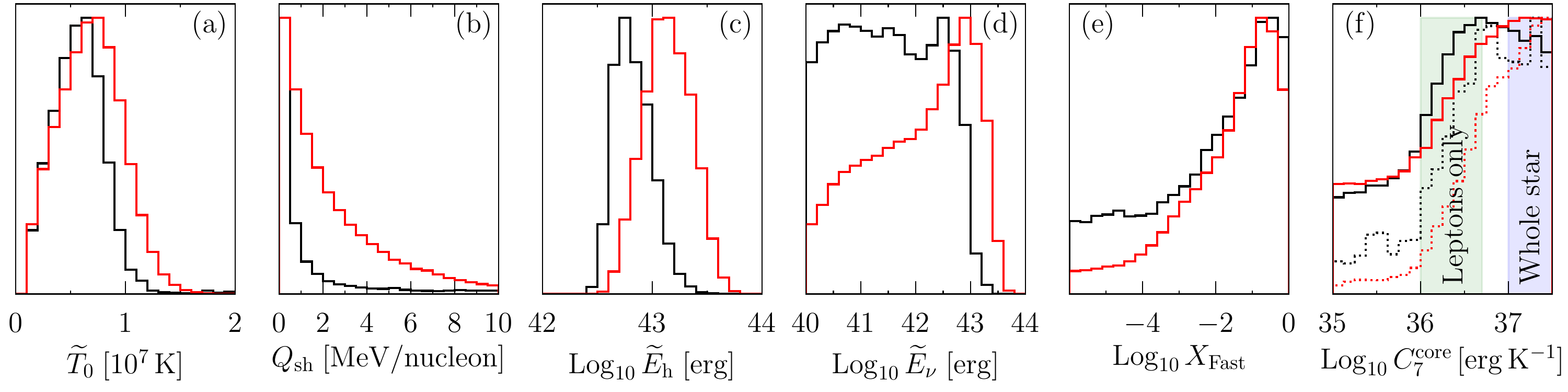}
\end{center}
\caption{Histograms of the distributions of our most important MCMC results:
(a) initial redshifted core temperature, $\widetilde{T}_0$, 
(b) strength of shallow heating, $Q_\mathrm{sh}$, 
(c) total redshifted heat injected into the star during the whole outburst, $\widetilde{E}_\mathrm{h}$,
(d) total redshifted energy lost to neutrinos during the outburst and till our 2018 observation, $\widetilde{E}_\nu$,
(e) neutrino luminosity scale factor $X_\mathrm{Fast}$, 
and
(f) core heat capacity rescaled at $\widetilde{T} = 10^7$ K, $C_7$
with the two range of ``leptons only'' and ``whole star'', as defined in Figure~\ref{fig:ALL1}, outlined.
Black and red curves correspond to our MCMC runs A and B, described in Section~\ref{sec:MCMC}
and detailed in Appendix~\ref{app:setupMCMC}.
In panel (f) dotted curves correspond to the subsets of results restricted by the condition 
$X_\mathrm{Fast} < 10^{-4}$,
i.e., models with no significant fast neutrino emission. 
Vertical scales are linear and chosen such that all curves have the same maximum.
}
\label{fig:FINAL}
\end{figure*}

\subsection{Results for \HETE}

We present in \fig{fig:FINAL} the most relevant results from our two MCMC runs. 
In spite of differences between our run A and B, corresponding to different priors on the thermal conductivity,
we see that \HETE's core temperature $\widetilde{T}_0$ (panel a) is very likely below $10^7$ K. For the shallow heating strength,
$Q_\mathrm{sh}$ (panel b), differences between runs A and B are more noticeable, which is also reflected by the noticeable differences in the total amount of heat injected into the star, $\widetilde{E}_\mathrm{h}$ (panel c), and 
the total amount of energy lost to neutrinos $\widetilde{E}_\nu$ (panel d). 
However, when considering our main results, i.e., constraints on the fast neutrino emission $X_\mathrm{Fast}$ and heat capacity $C^\mathrm{core}$ in panels (e) and (f), differences between runs A and B are minor. 
We see in panel (e) that there is a strong preference for significant fast neutrino emission. We note that a very low neutrino emission rate $\widetilde{L}_\mathrm{Fast}$, 
i.e. with $X_\mathrm{Fast} < 10^{-4}$, cannot be excluded but results in models with $\tau_\mathrm{rec}$ of several thousands of years, which may not be very likely (see Section~\ref{sec:nu}).

For the core heat capacity, panel (f) of \fig{fig:FINAL}, we find that its value at $\widetilde{T} = 10^7$ K,
$C^\mathrm{core}_7$, is preferred to be above $10^{36}$ erg K$^{-1}$. 
However, there is also a significant tail below $10^{36}$ erg K$^{-1}$ that comprises just 
slightly less than a quarter of the total posterior.
Most interesting are models with $C^\mathrm{core}_7$ above the maximum lepton contribution of 
$5 \times 10^{36}$ erg K$^{-1}$ (see \eq{Eq:Clep})
since this implies that some fraction of the baryons are {\it not} paired:
we find that about 43\% of our models in Scenario A, and 46\% in Scenario B, are in this upper range.
If we restrict ourselves to models that exclude extreme phases of matter 
(such as CFL quark matter) and that have $C_7$ larger than the minimal value from only the leptons, 
$10^{36}$ erg K$^{-1}$, 
then the fraction of models with $C^\mathrm{core}_7$ above $5 \times 10^{36}$ erg K$^{-1}$ that imply 
some fraction of baryon not being paired is now 55\% in Scenario A and 60\% in Scenario B.

\begin{figure}
\begin{center}
\includegraphics[width=0.99\columnwidth]{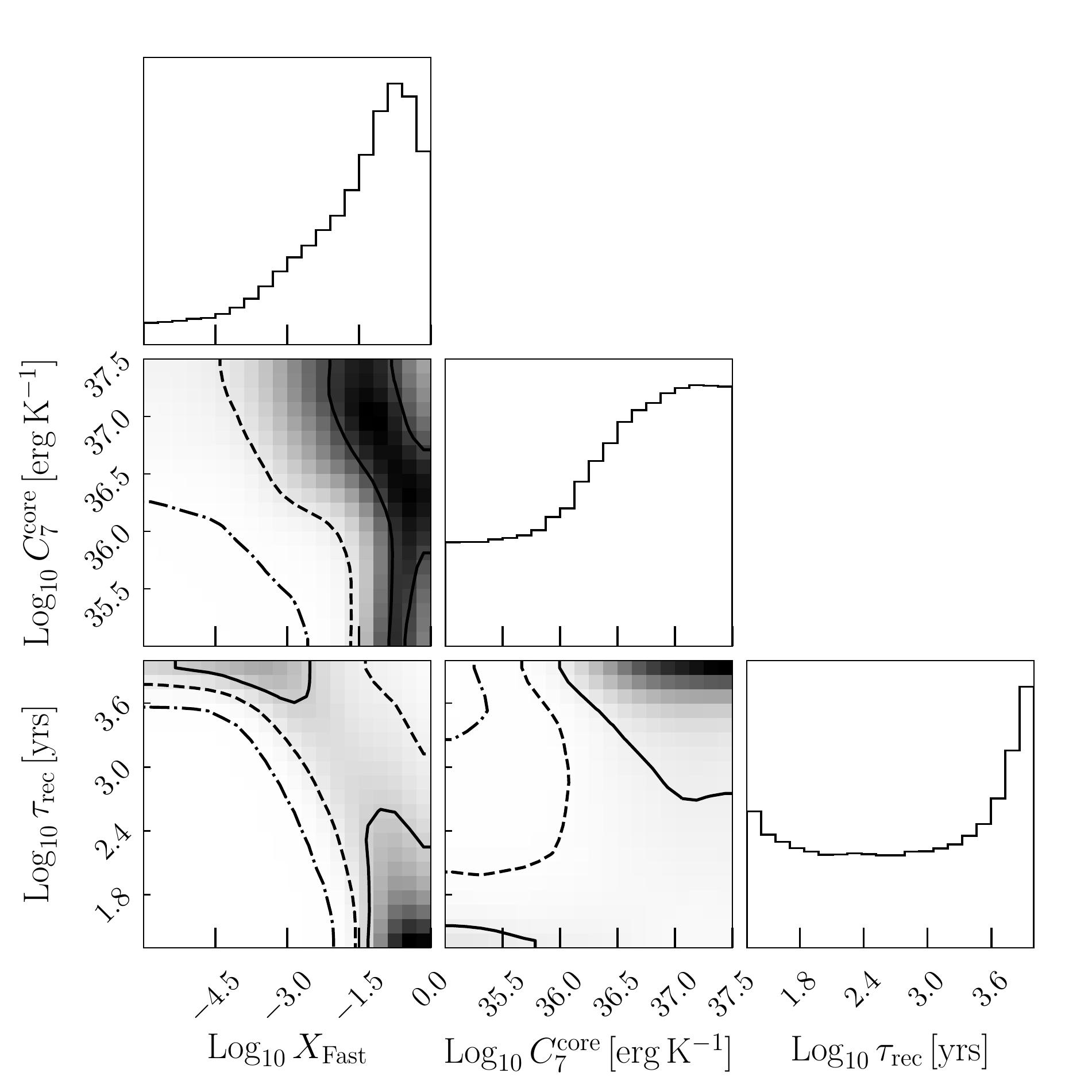}
\end{center}
\caption{Correlation between the recurrence time, fast neutrino emission (Equation~\ref{Eq:L_Fast1}),
and core heat capacity (Equation~\ref{Eq:C_core}) from our MCMC run B.
(Our MCMC run A gives very similar results.)
Contour lines, continuous, dotted, and dash-dotted, show 1, 2, and $3\sigma$ confidence ranges, respectively.
}
\label{fig:CXF}
\end{figure}

The dotted curves in panel (f) illustrate that the weak constraint on $C_7$ is directly due to the possibility of fast neutrino emission: restricting ourselves to models with $X_\mathrm{Fast} < -4$, i.e., models where $L_\mathrm{Fast}$ is just comparable to or even lower than $L_\gamma$, we would have a strong constraint with 95\% in run A and 87\% in run B posterior probability of $C^\mathrm{core}_7 > 36$, i.e., implying a nucleon contribution to the specific heat. 
\fig{fig:CXF} further illustrates that the low $C^\mathrm{core}$ regime is directly correlated with high neutrino emission and also exhibits the clear correlation between the recurrence time $\tau_\mathrm{rec}$ and the fast neutrino emission.

We also add in \fig{fig:ALL1} contour lines, at $1\sigma$, $2\sigma$, and $3\sigma$, from our MCMC run B.
Uncertainties on $M$, $R$, and $D$ affect the deduced value of $T_\mathrm{eff}^\infty$ in 2018 (see \fig{fig:grid})
and result in a broader range of inferred values of $\widetilde{T}_1$.
Similarly, our MCMC results imply a broader range of estimated deposited energy, $\widetilde{E}_\mathrm{h}$,
due to both the uncertainty in $D$ that directly impacts on the inferred $\dot{M}$ and the uncertainty on the shallow heating.
Adding the possibility of fast neutrino cooling, dotted contour lines in this same \fig{fig:ALL1}, significantly
increases the range of the inferred $\widetilde{T}_1$ and total heat capacity as also presented in 
Figures \ref{fig:FINAL} and \ref{fig:CXF}.

Having presented our main results here, we briefly describe in Appendix~\ref{app:results} two additional results of our MCMC runs that provide some complementary analysis on our main results: we discuss how our simulations seem to prefer a small NS radius and large mass and concisely prove that crustal matter forms a solid lattice.

\section{Conclusions}
\label{sec:conclusions}
We presented a new post-accretion outburst \chan\ observation of \HETE, obtained in 2018. While only 6 net photons were detected in our new observations, we performed extensive simulations to infer a temperature of $T_{\mathrm{eff}}^{\infty}\approx 30-39$~eV, depending on the assumed distance, mass, and radius. The new data thus reveal that the NS experienced significant further cooling compared to our previous observation obtained in 2016, during which we measured $T_{\mathrm{eff}}^{\infty}\approx 55$~eV \citep[for $D = 4.7$~kpc, $M = 1.4~\Msun$, and $R = 10$~km;][]{degenaar2017}.

For our new observation, we infer a thermal photon luminosity of $L_\mathrm{bol} \approx 10^{31}$ erg s$^{-1}$, which makes \HETE\ one of the three dimmest quiescent NSs in an LMXB. The other two are \sax, which has a thermal luminosity $L_\mathrm{bol} < 6 \times 10^{30}$ erg s$^{-1}$ \citep{2009ApJ...691.1035H}\footnote{We note that \sax\ is detected in quiescence with an X-ray luminosity of $L_\mathrm{X} \approx 10^{32}$~erg~s$^{-1}$ (0.5--10 keV), but is fully dominated by power-law emission (of unknown origin) and does not appear to contain any significant surface emission from the NS. Modelling of the X-ray data provides an upper limit on such thermal emission as quoted in the text \citep{2009ApJ...691.1035H}.}, and 1H~1905+000 with $L_\mathrm{bol} < 2.4 \times 10^{30}$ erg s$^{-1}$ \citep{2007ApJ...665L.147J}. 

Very cold NSs that are heated during accretion outbursts in LMXBs can potentially pose interesting constraints on the properties of the ultra-dense matter in their cores \citep[e.g.,][]{cumming2017,brown2018}. Even though it is generally assumed that an NS in a binary system is heated during an accretion phase, there are no direct observational constraints showing that this happens in \sax, 
nor are there data to show this for 1H~1905+000. However, in the case of \HETE, the two \chan\ observations of 2016 and 2018 clearly show significant cooling implying that, in \HETE, heating of the NS crust did occur. 
We set out to see if the new, lower temperature measurement of \HETE\ can provide improved constraints on the heat capacity of the NS core. To this end, we performed simulations of its thermal evolution during and after the accretion phase, taking into account many physical and astrophysical uncertainties (as detailed in Section~\ref{sec:MCMC} and Appendix~\ref{app:setupMCMC}). Interestingly, we find probable solutions that require a high heat capacity, suggesting that nucleons have a significant contribution to the heat capacity (i.e., cannot be paired in a superfluid). However, the strong probability of \HETE\ exhibiting fast neutrino cooling prohibits us from drawing strong conclusions.

Simple arguments suggest that the NS in \HETE\ is likely exhibiting rapid core neutrino cooling. The $\approx$10-yr long accretion outburst of \HETE\ was very well covered by all-sky X-ray monitors, and hence we have a reasonable handle on the amount of heat that was injected into the NS. Combining this with the very low observed core temperature would require the system to have a recurrence time of thousands of years. This would appear unreasonable since many Galactic NS LMXBs have relatively high duty cycles on the order of $\sim$10\% \citep[e.g.,][]{degenaar2010,yan2015}. However, selection effects could limit our verification of LMXBs with very low duty cycles \citep[e.g.,][]{heinke2010_GC,wijnands2013}. For recurrence times that are not this excessively long, explaining the very low core temperature would require the NS to exhibit rapid neutrino cooling. Indeed, our thermal evolution simulations provide solutions with very fast neutrino emissions. These solutions, however, allow for a low heat capacity that can simply be explained by leptons.

On theoretical grounds, a high neutrino luminosity requires a large number of thermally excited pseudo-particles\footnote{In many-body theory of strongly interacting particles, this term is used to indicate that a particle is excited and dragging/pushing other particles around as if moving with a cloud of excitations around it.} that can 
participate in the corresponding process and these very same pseudo-particles do provide a significant contribution to the heat capacity. Quantifying this contribution is, however, very model dependent. 
As seen in \fig{fig:CXF}, low values of $C^\mathrm{core}$ favor $X_\mathrm{Fast} \sim 0.1$.
A value of $X_\mathrm{Fast} \sim 0.1$ could be provided by a nucleonic direct-Urca process acting in about 10\% of the core's volume implying that at least 10\% of the core nucleons are {\it not} paired and that $C^\mathrm{core}_7 > 10^{36}$ erg K$^{-1}$.
It could also, however, be provided by a meson condensate, whose neutrino emissivity is much lower, which would then imply that
most of the core's volume contribute to this emission and thus, most of the core's nucleons are not paired resulting in a very large heat capacity, $C^\mathrm{core}_7 \gg 10^{36}$ erg K$^{-1}$.
These constraints are, however, strongly based on theoretical arguments and cannot be considered as constraints
directly derived from the data.

We end up with the interesting situation that our modeling of \HETE\ requires the NS core to have either a very high heat capacity or a very high rate of neutrino emission. Both options suggest that a significant fraction of the core particles cannot be paired in a superfluid. Our thermal evolution simulations of \HETE\ predict that the NS may not have had sufficient time to thermally relax from its accretion phase. We find that further cooling, possibly to temperatures as low as $\approx15$~eV, may have occurred after our 2018 measurement. A new, deeper \chan\ observation of \HETE\ may provide more stringent constraints on its core temperature. This brings about the exciting prospect that it may allow us to set a tighter limit on the fraction of baryons that is paired.

\section*{Acknowledgements}
ND is supported by a Vidi grant from the Netherlands organization for scientific research (NWO). DP and MB acknowledge financial support by the Mexican Consejo Nacional de Ciencia y Tecnolog{\'\i}a with a CB-2014-1 grant $\#$240512 and the Universidad Nacional Aut\'onoma de M\'exico through an UNAM-PAPIIT grant \#109520. MB also acknowledges support from a postdoctoral fellowship from UNAM-DGAPA. JvdE is  supported  by  a  Lee  Hysan  Junior  Research  Fellowship from St Hilda’s College, Oxford. MR acknowledges support from CXO grant GO8-19031X. This work made use of results provided by the \rxte/ASM team, \maxi\ data provided by RIKEN, JAXA and the \maxi\ team, and \swift/BAT transient monitor results provided by the \swift/BAT team. The authors are grateful to the referee, Craig Heinke, for useful comments.

\section*{Data Availability Statement}
The data underlying this article are available in Zenodo, at DOI 10.5281/zenodo.4488241. These astrophysical data sets were derived from sources in the public domain: https://cda.harvard.edu/chaser/ and https://heasarc.gsfc.nasa.gov/cgi-bin/W3Browse/swift.pl.

\bibliographystyle{mn2e}


\appendix

\section{Setup of our MCMC runs}
\label{app:setupMCMC}

\begin{figure*}
\centerline{ \includegraphics[width=1.0\textwidth]{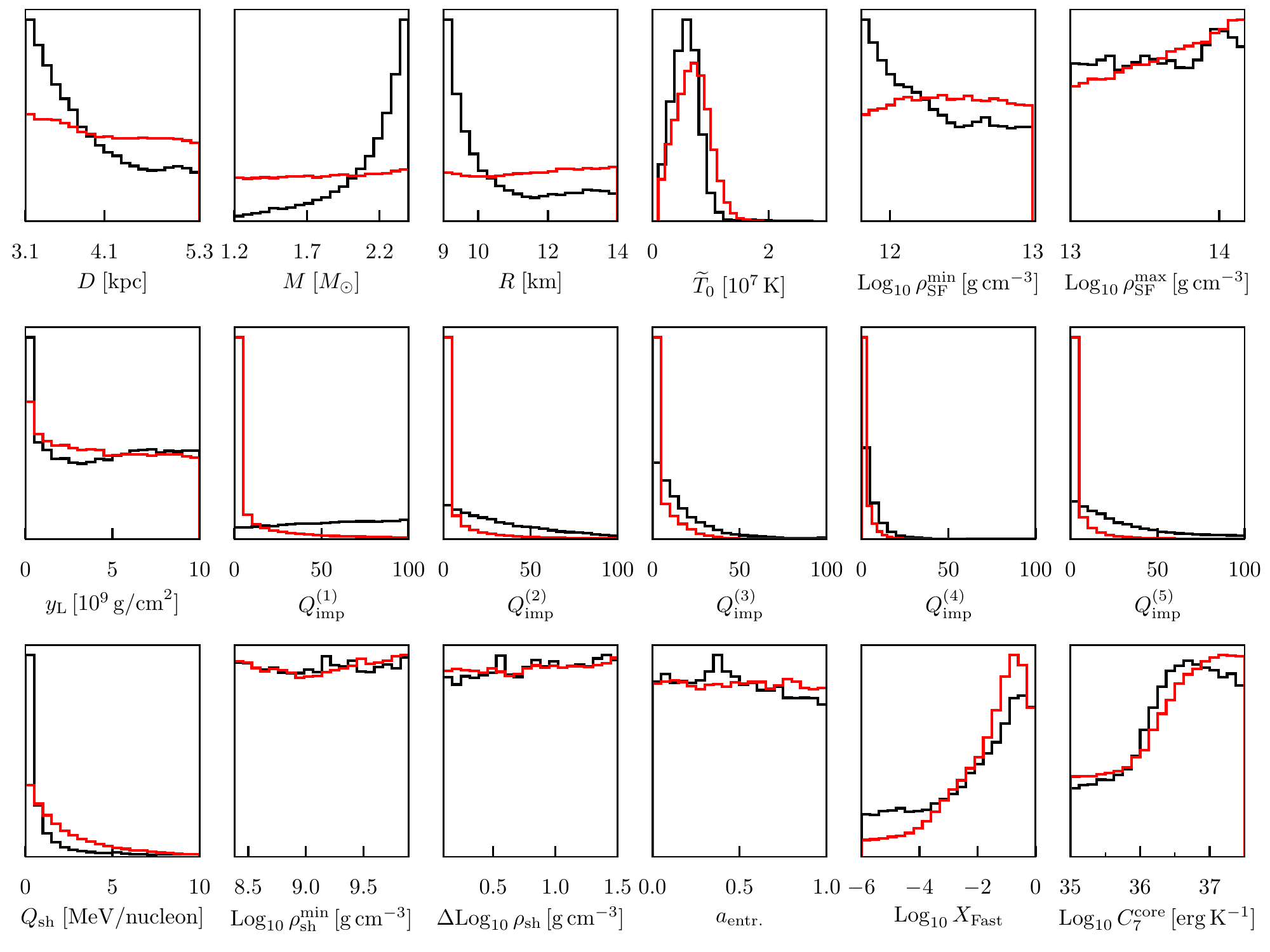} }
\caption{Histograms of the posterior distributions of our 18 MCMC parameters; see text for their description.
Black and red lines correspond to our MCMC runs A and B, respectively and
vertical scales are arbitrary but histograms are renormalized so that both have the same area.
Priors were linear and uniform over the displayed range except for the $Q_\mathrm{imp}^{(i=1, \dots, 5)}$ in run B where the prior was logarithmic, i.e.,
$\mathrm{Log}_{10} Q_\mathrm{imp}^{(i=1, \dots 5)}$ uniformly distributed over the range of $-2$ to $+2$. Vertical scales are linear and chosen such that all curves have the same maximum.
}
\label{fig:histograms_all}
\end{figure*}

We employ the Markov-chain Monte Carlo techniques (see e.g., \citealt{Gregory:2005aa}) to cover extensively the parameter space of both astrophysical and
physical uncertainties in our understanding of the \HETE\ structure and evolution.
In total, we have 18 parameters, which we describe below.
These cover, and probably exaggerate, the range of uncertainties on the microphysics and the astrophysical settings and give us confidence that
our results will be as model-independent as possible.
The posterior distributions of our parameters are shown in \fig{fig:histograms_all} and their prior distributions were uniform in the range shown in the figure,
either linear or logarithmic as labelled, with the exception of the five $Q_\mathrm{imp}^{(i=1, \dots, 5)}$ in run B for which the prior was logarithmic, i.e., linear in 
$\mathrm{Log}_{10 }Q_\mathrm{imp}^{(i)}$ with a range from $-2$ to $+2$.

The first three parameters are the distance, $D$,  the mass, $M$, and the radius, $R$, of the NS.
These three properties impact the data interpretation as described in \sect{sec:obs} and displayed in \fig{fig:grid}.
Moreover, $M$ and $R$ determine the thickness of the NS crust, the essential region where the time evolution constrained by our two temperature
measurements is taking place.
Importantly too, $D$ controls our inference of the mass accretion rate $\dot M$ which is $\propto D^2$.
The prior range of $D$ is taken from \cite{Galloway:2008aa} while the range of $M$, 1.2--2.4 $M_\odot$, is chosen to cover the range of presently known pulsar masses 
\citep{Lattimer:2012aa,Ozel:2016aa}.
For the radius range we consider both the NICER \citep{miller2019,riley2019} and the LIGO \citep{abbott2018_NSmerger_EOS} results and assume $R$ to be between 9 and 14 km.

The fourth parameter is the initial redshifted (uniform) internal temperature of the NS, $\widetilde{T}_0$.
Since \HETE\ was discovered at the initiation of its only known outburst, we have no information about its pre-outburst thermal state 
and thus take $\widetilde{T}_0$ as a free parameter within a range from almost zero up to $3\times 10^7$ K. 
The posterior distribution of $\widetilde{T}_0$ shows that this range covered all possible values. 

Next, the five  parameters 5--9 control the thermal conductivity in the crust through the {\it impurity parameter} $Q_\mathrm{imp}$
that takes different values in five different zones.
At the low temperatures present in the crust of \HETE, the thermal conductivity is strongly dominated by electrons.
Electron scattering is controlled by ion scattering when these are in a liquid phase, in which case we follow 
\citet{Yakovlev:1980aa}, and by phonon scattering when ions are crystalized, where we employ the results of 
\citet{Gnedin:2001aa} augmented by impurity scattering. 
For the latter we apply a simple formalism
(\citealt{Flowers:1976aa,Yakovlev:1980aa,Itoh:1993aa}, but see also \citealt{Roggero:2016aa})
in which the electron-impurity scattering rate is written as 
$\nu_\mathrm{e-imp} = Q_\mathrm{imp} \times \nu_\mathrm{e-imp}^{(1)}$ where $\nu_\mathrm{e-imp}^{(1)}$
is a fiducial frequency.
Formally defined as the average-squared charge fluctuation, $Q_\mathrm{imp} = \langle (Z - \langle Z \rangle)^2 \rangle$ 
where $Z$ is the charge number of nuclei and $\langle Z \rangle$ its average, 
modelling of X-ray bursts and the evolution of nuclei in the accretion-compressed crust indicate that large values
are possible \citep{Gupta:2007aa,Gupta:2008aa}, while phase separation at crystallization is predicted to reduce it \citep{Horowitz:2009aa}.
In modelling crust relaxation in LMXBs $Q_\mathrm{imp}$ has been treated as a free parameter, as we do here, and small values near unity have usually been preferred
when fitting observations (see e.g., \citealt{Brown:2009aa,pagereddy2013,Turlione:2015aa,Ootes:2018aa,Parikh:2019aa}), but some large values have also been reported \citep[$Q_\mathrm{imp} \approx 40$;][]{Degenaar:2014aa}.
Moreover, $Q_\mathrm{imp}$ is not expected to be uniform within the crust, so we split the latter in five zones,
employing values $Q_\mathrm{imp}^{(i=1, \dots, 5)}$ in the density range $\rho^{(i-1)} - \rho^{(i)}$ with $\rho^{(i)} \equiv 10^{10+i}$ g cm$^{-3}$.

The tenth model parameter reflects the chemical composition of the NS envelope, which is important because it determines the ``$T_\mathrm{b} - T_\mathrm{eff}$'' relationship, i.e. the relationship between
the boundary temperature $T_\mathrm{b}$ at the last zone of the outer crust included in the numerical simulation at density $\rho_\mathrm{b} = 10^8$ gm cm$^{-3}$
and the surface effective temperature, $T_\mathrm{eff}$.
It is a potentially important, and essentially free, parameter. 
The presence of light elements, as H, He, or C deposited by the accretion and/or produced by surface nuclear burning, is naturally expected and
results in an increase in $T_\mathrm{eff}$ for a given $T_\mathrm{b}$ compared to a model composition comprising heavy elements as Fe.
This is parametrized by the column density of light elements $y_\mathrm{L}$ and our envelope model was displayed in figure~3 of \citet{degenaar2017}. 
We note that this figure shows that the mapping between the observed surface temperature and the interior temperature is virtually independent of the (uncertain) light-element content of the envelope
for the very low temperature of \HETE\ found in our 2018 observation but not in the case of the 2016 observation.

Parameters 11--13 then characterize the shallow heating in the crust, with parameter 11 defining its strength and 12--13 the density range in the crust over which it is acting. As mentioned in the main body of the article, a major unknown in modelling crust relaxation in NS-LMXBs has turned out to be the presence of this shallow heating: its nature, and hence what determines its magnitude, is still an open issue.
We describe it numerically by a time-dependent energy injection, per unit volume, $q_\mathrm{sh}(t)$, as
\begin{equation}
q_\mathrm{sh}(t) = \frac{Q_\mathrm{sh}}{\Delta V} \frac{\dot{M}}{m_u}
\label{Eq:q_sh}
\end{equation}
which acts within a shell of volume $\Delta V$ delimited by the densities $\rho_\mathrm{sh}^\mathrm{min}$
and $\rho_\mathrm{sh}^\mathrm{max} = \rho_\mathrm{sh}^\mathrm{min} + \Delta \rho_\mathrm{sh}$
so that a total energy $Q_\mathrm{sh}$ is injected per accreted baryon ($m_u$ being the atomic mass unit).
$\rho_\mathrm{sh}^\mathrm{min}$, $\Delta \rho_\mathrm{sh}$, and $Q_\mathrm{sh}$ are our MCMC parameters.
As maximum strength, we consider 10 MeV per nucleon, the strong value found in MAXI J0556-532 \citep{Deibel:2015aa,Parikh:2017aa}.

Next, parameters 14--16 describe the occurrence of neutron superfluidity in the crust. The occurrence of superfluidity affects the crust specific heat, and the major uncertainty here is the state of the dripped neutrons in the inner crust. If these are normal, they have an enormous contribution that dominates all other components (electrons and phonons), while when in a superfluid state their contribution becomes largely negligible. Theoretical predictions for the density dependence of the neutron singlet superfluid critical temperature $T_c$ show bell-shaped curves with a maximum at mid densities
($\approx 10^{13}$ g cm$^{-3}$) that vanishes around nuclear matter density. Typical maximum values are of the order of $10^{10}$ K, higher than temperatures present in \HETE\ by about 3 orders of magnitude. 
The critical regions where uncertainties on the shape of the $T_c(\rho)$  are of importance to determine the extent of the normal/superfluid regimes are thus the two regions just above neutron drip where $T_c(\rho)$ grows rapidly and when approaching the crust-core interface where $T_c(\rho)$
decreases and then vanishes.
To incorporate these uncertainties in a simple fashion, we introduce two parameters (numbers 14 and 15),  $\rho_\mathrm{SF}^\mathrm{min}$ and $\rho_\mathrm{SF}^\mathrm{max}$, and posit that neutrons are superfluid with $T_c \approx 10^{10}$ K in between these two densities and normal below $\rho_\mathrm{SF}^\mathrm{min}$ and above $\rho_\mathrm{SF}^\mathrm{max}$.

Dripped neutrons can moreover be entrained by the nuclei in their vibrational motion because of the coupling to the lattice through band effects \citep{Chamel:2005aa}.
The main effect of entrainment, in our present concerns, is to increase the effective mass of nuclei in the inner crust, thus reducing the speed of phonons, 
transverse as well as longitudinal modes, which results in an increase in the lattice specific heat when cold enough to be in the Debye regime \citep{Chamel:2013aa}.
This effect is parametrized by the fraction $a_\mathrm{entr}$ of entrained dripped neutrons (model parameter 16).

The penultimate parameter $X_\mathrm{Fast}$ controls the fast neutrino emission whose luminosity is given by \eq{Eq:L_Fast1}.
We implement this energy loss in the inner half of the core (since it is theoretically expected to act only at high densities)
with an emissivity $q_\nu^\mathrm{Fast}(T) = q_0 \times T^6$, $q_0$ being a constant independent of density or temperature
such that the integral of $q_\nu^\mathrm{Fast}(T)$ at uniform $\widetilde{T} = 10^7$ K gives the luminosity of 
Equation~\ref{Eq:L_Fast1}.
At the low core temperature of \HETE\ the thermal conductivity is so large that the core always remains essentially isothermal, even during the accretion outburst,
and so how the fast neutrino emission is actually distributed within the core is inconsequential.

Our last parameter is the star core's heat capacity, $C^\mathrm{core}$, as defined in Equation~\ref{Eq:C_core}.
As in the case of fast neutrino emission, for the numerical calculation with \textsc{NSCool}, we need the specific heat 
per unit volume, which we write as $c(T) = c_0 \times T$, $c_0$ being a constant such that when $c(T)$ is integrated 
over the whole core at a uniform $\widetilde{T} = 10^7$ K, it gives the total heat capacity of Equation~\ref{Eq:C_core}.

\section{Two More Results of our MCMC runs}
\label{app:results}

We provide here additional results of our MCMC runs that were not described in the main text and give some complementary analysis of the main results that were. 
This concerns hints for a preference of small NS radius and large mass and concise proof that crustal matter forms a solid lattice.

As a first point we consider the main difference between run A, with linear prior, and run B, with logarithmic prior on $Q_\mathrm{imp}^{(i=1, \dots, 5)}$.
The latter gives more preference to low values of $Q_\mathrm{imp}^{(i)}$ and its effect is clearly seen in the posteriors in the corresponding panels of \fig{fig:histograms_all}.
As a result, in run B, preferentially high crust thermal conductivities (because of preferentially low impurity contents) make it easy for models to have short crustal
thermal relaxation times needed for the cooling from the observed 2016 $T_\mathrm{eff}^\infty$ to the 2018 one.
In contradistinction, in run A, preferentially high $Q_\mathrm{imp}$ and lower conductivities require more, so that the MCMC finds, in such circumstances, a preference for  high masses $M$ and small radii $R$ as seen in the corresponding panels of \fig{fig:histograms_all}, while no such preference is found in run B.\footnote{High mass and small radius imply very thin crust allowing a short relaxation time even with not so high thermal conductivity.} This finding from run A, high mass and small radius, is very interesting in that it may be supported by the preferences, in both runs A and B, of very efficient neutrino emission
(i.e., large $X_\mathrm{Fast}$), which is more likely to occur in a very massive star that also may naturally have a small radius.
However, large $X_\mathrm{Fast}$ is not a strong result as small values are far from excluded, in both runs A and B, and preference for large $M$ with small $R$
depends on the assumed prior for $Q_\mathrm{imp}$. Therefore, we can only leave this result as a suggestion which calls for further inquiries.

Finally, as a curiosity, notice that $Q_\mathrm{imp}$ in the zones 3 and 4, i.e., the density ranges $10^{12}$ - $10^{13}$ and $10^{13}$ - $10^{14}$ g cm$^{-3}$, 
respectively, is strongly restricted to be less than 50. 
This implies that, in this density range, {\it crustal matter forms a solid with a lattice structure} in contradistinction with an amorphous solid
that would have a much lower thermal conductivity.
This significant constraint results directly from the imposition of crust cooling between the 2016 and 2018 observations resulting from a drop in
$T_\mathrm{eff}^\infty$ from $\approx 50$ down to $\approx 35$ eV in 2 yr.
Impressively, with two data points and six photons for the second one, we can conclude that the NS crust forms a crystal. This result had already been obtained in essentially all previous studies of crust cooling in NS-LMXBs but \HETE\ provides the most concise (six photons) proof.

\end{document}